\documentclass[aps,pre,twocolumn,groupedaddress,superscriptaddress,showpacs]{revtex4-1}
\usepackage{amsmath}
\usepackage{graphicx}
\usepackage{float}
\usepackage{xcolor}

\begin{document}

  \title {Nonequilibrium growth of patchy-colloid networks on substrates}

  \author{C. S. Dias}
   \email{csdias@cii.fc.ul.pt}
    \affiliation{Departamento de F\'{\i}sica, Faculdade de Ci\^{e}ncias da Universidade de Lisboa, P-1749-016 Lisboa, Portugal and Centro de F\'isica Te\'orica e Computacional, Universidade de Lisboa, Avenida Professor Gama Pinto 2, P-1649-003 Lisboa, Portugal}

  \author{N. A. M. Ara\'ujo}
   \email{nuno@ethz.ch}
   \affiliation{Computational Physics for Engineering Materials, IfB, ETH Zurich, Wolfgang-Pauli-Strasse 27, CH-8093 Zurich, Switzerland}

  \author{M. M. Telo da Gama}
   \email{margarid@cii.fc.ul.pt}
    \affiliation{Departamento de F\'{\i}sica, Faculdade de Ci\^{e}ncias da Universidade de Lisboa, P-1749-016 Lisboa, Portugal and Centro de F\'isica Te\'orica e Computacional, Universidade de Lisboa, Avenida Professor Gama Pinto 2, P-1649-003 Lisboa, Portugal}


  \begin{abstract}
Patchy colloids with highly directional interactions are ideal building
blocks to control the local arrangements resulting from their
spontaneous self-organization. Here we propose their use, combined with
substrates and nonequilibrium conditions, to obtain novel structures,
different from those of equilibrium thermodynamic phases. Specifically,
we investigate numerically the irreversible adhesion of three-patch
colloids near attractive substrates, and analyze the fractal network of
connected particles that is formed. The network density profile exhibits
three distinct regimes, with different structural and scaling
properties, which we characterize in detail. The adsorption of a mixture
of three- and two-patch colloids is also considered. An optimal fraction
of two-patch colloids is found where the total density of the film is
maximized, in contrast to the equilibrium gel structures where a
monotonic decrease of the density has been reported. 
  \end{abstract}

  \maketitle

\section{Introduction}
The past few years have witnessed a sustained interest in the
self-organization of patchy colloids, with the development of a
wide range of techniques to synthesize them
\cite{Shum2010,Duguet2011,Sacanna2011,He2012,Hu2012,Wilner2012}. These
particles, with functionalized surfaces, yield new features such as
anisotropic interactions, control of the valence, and the formation of
permanent electrical dipoles, paving the way to the development of novel
materials with fine tuned mechanical, optical, and thermal properties
\cite{Pawar2010, Kretzschmar2011, Sacanna2011, Bianchi2011}.
Understanding how nonequilibrium conditions influence this
self-organization is crucial to develop strategies to design new
materials, as the novel structures emerge at very low temperatures,
where thermal and mechanical equilibration might be difficult to achieve
under normal experimental conditions. 

Theoretical and experimental studies of patchy colloids have been
focused on their cooperative behavior in solution \cite{Russo2009,
Sciortino2011, Sciortino2007, Bianchi2011}, where several models were
considered with the aim of describing a range of more complex building
blocks such as amphiphilic molecules, colloidal clays, proteins, and DNA
nano-assemblies \cite{Glotzer2004, Doye2007, Sciortino2010, Pawar2010,
Glotzer2010, Bianchi2011, Sciortino2011, Rosenthal2011, Ruzicka2011}. By
contrast, the investigation of self-organization of the simplest of
these models at planar substrates is only just beginning \cite{Gnan2012,
Bernardino2012}.  Theoretical studies have shown that, in the presence
of a substrate, a rich equilibrium phase diagram emerges with very
unusual properties such as, e.g., two wetting transitions and a
nonmonotonic surface tension \cite{Bernardino2012}. These works have
been restricted to equilibrium features; however, it is widely
recognized that the use of substrates might improve the degree of
control of aggregation, especially under nonequilibrium conditions
\cite{Cadilhe2007, Araujo2008}. For example, a growth direction can be
defined which allows the use of kinetic features to control the film
structure \cite{Einstein2010}.

\begin{figure}[t]
  \begin{center}
  \includegraphics[width=0.7\columnwidth]{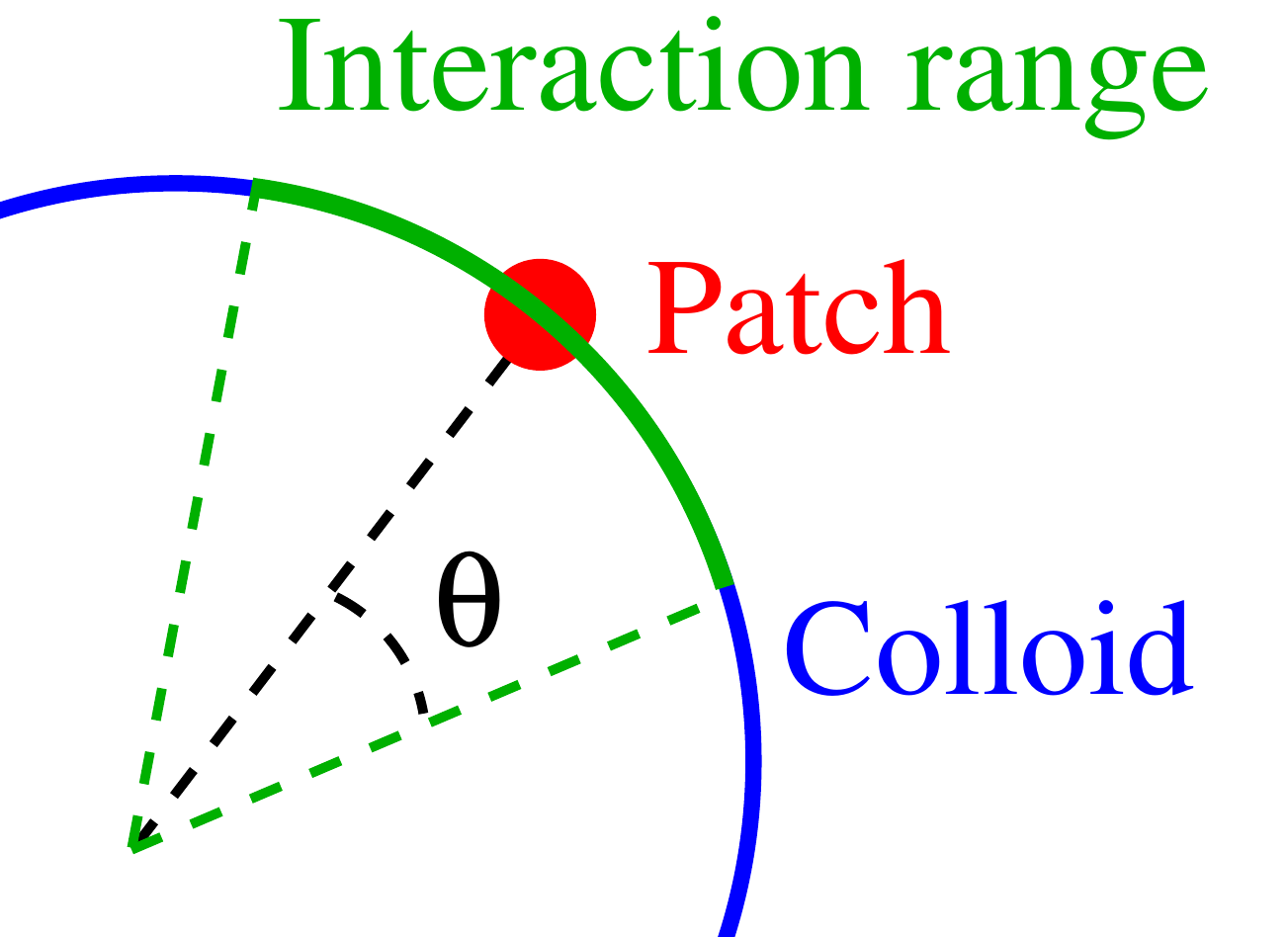} \\
  \end{center}
  \caption{
  (color online) Schematic representation of a patch (red) on the surface of a
colloid (blue) and its interaction range (green). The limits of the
interaction range are defined by an angle $\theta$ with the center of
the patch.
   \label{fig.figure1}
   }
\end{figure}

In this paper we address the nonequilibrium adsorption of patchy
colloids on substrates, resulting from the irreversible nature of the
binding, and we characterize the resulting network of
connected particles. In deep contrast with equilibrium films, a fractal
network is assembled with a fractal dimension compatible with the one
reported for Diffusion Limited Aggregation (DLA) \cite{Meakin1998,
Witten1981}. We systematically analyze the dependence of the network
structure on the substrate size and diffusion coefficient of the
colloids in solution (which can be controlled experimentally, for
example, by the thermostat temperature). We show that, although the
density of the film strongly depends on the diffusion coefficient, the
fractal dimension is resilient over a wide range of growth conditions.
By contrast with previous models based on DLA, here directional
interactions are considered with results that depend on the diffusion
coefficient and colloidal valence.

It has been shown that the distribution of patches affects the
aggregation process and, consequently, the equilibrium bulk structures
\cite{Sciortino2007, Bianchi2007, Bianchi2008, Russo2009, Ruzicka2011,
Kretzschmar2011}. For example, control of the valence allows tuning up
the density and temperature of both the gas-liquid and sol-gel critical
points \cite{Bianchi2006, Russo2009, Russo2010}. Here, we consider
three-patch colloids and investigate the density profile of the adsorbed
nonequilibrium network. We then proceed to investigate the adsorption of
three- and two-patch colloids, which can be synthesized with the
available experimental techniques
\cite{Cho2005,Blaaderen2006,Pawar2010,Duguet2011,Kraft2012,Wang2012},
and characterize the dependence of the film density on the concentration
of the mixture.  Studies of the coexisting thermodynamic structures of
mixtures of two- and three-patch colloids in solution reveal a monotonic
decrease of the density with the fraction of two-patch colloids
\cite{Bianchi2006, Bianchi2008, Liu2009}. Somewhat surprisingly, we have
found that the density of the adsorbed network film, under
nonequilibrium conditions, increases with the concentration of two-patch
particles before it decreases, exhibiting a well defined maximum at an
intermediate concentration.   

In the following section we describe the model. 
In Sec.~\ref{sec.results}, we quantify these results and illustrate the full
scaling behavior of the nonequilibrium adsorbed film. 
Finally, in Sec.~\ref{sec.conclusions}, we draw some conclusions.

\section{Model}\label{sec.model}

\begin{figure}[t]
  \includegraphics[width=\columnwidth]{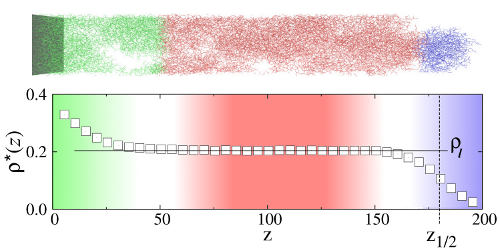} \\
  \caption{(color online) Density profile of the colloidal network on a
substrate showing three different regimes: surface layer (green, left);
liquid film (red, middle); interfacial region (blue, right).
\textit{Top:} Snapshot of a typical configuration, where each stick
corresponds to a colloid-colloid connection. \textit{Bottom:} Density,
$\rho^*(z)$, as a function of the height, $z$, after the adsorption of
$40$ particle layers, averaged over $500$ independent realizations.
$\rho_l$ is the density of the liquid film and $z_{1/2}$ is the film
thickness, defined as the height at which the density is $\rho_l/2$.
  \label{fig.figure2}}
\end{figure}

In order to simulate the colloidal adsorption we propose and use a stochastic model
based on the nonequilibrium Monte Carlo (MC) method. Patchy colloids are
frequently described as spherical particles with a short-range repulsive
core and patch-patch attractive interactions. The attraction is
truncated at a certain angle around the center of the patch \cite{Villar2009}.

As show in Fig.~\ref{fig.figure1}, to account for particle-particle interaction we
define an interaction range (green), on the surface of the colloid
(blue), around each patch (red). This range is characterized by a single
parameter, namely, the angle $\theta$ with the center of the patch. Here
we have used $\theta=\pi/6$. In the event of a collision with a
pre-adsorbed colloid, if the contact point is within the interaction
range of the pre-adsorbed particle, the binding is successful with
probability $p$, corresponding to the fraction of the surface of the
landing colloid covered by the interaction range of all patches. In the
case of successful binding, the position of the landing particle is
adjusted based on the patch-patch orientation, otherwise, an elastic
collision occurs. The interaction range accounts for both the extension
of the patch on the colloid surface and the range of the patch-patch
interaction. In particular, the range of the patch-patch interaction is
 $\pi\sigma/6$, where $\sigma$ is the colloid diameter. Thus, our model
contains a single control parameter (the interaction range) that can be
tuned to mimic different patch-patch potentials.

\begin{figure}[t]
  \includegraphics[width=\columnwidth]{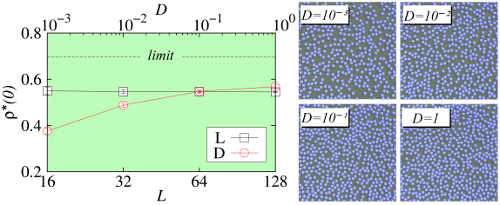}
  \caption{
(color online) \textit{Right:} Dependence of the density of
the first adsorbed layer, $\rho^*(0)$, on $D$, for values ranging
from $10^{-3}$ to $1$, and substrates with lateral size, $L$,
ranging from $16$ to $128$. \textit{Right:} Snapshots of the first
adsorbed layer (directly in contact with the substrate) for different
diffusion coefficients, $D$, namely, $10^{-3}$, $10^{-2}$, $10^{-1}$,
and $1$, on a substrate with lateral size 32 in units of the particle
diameter.
  \label{fig.figure3}}
\end{figure}

In the presence of a substrate, two characteristic timescales can be
identified: one related to the flux of colloids towards the substrate
(inter-arrival time) and the other to the binding between patches
(binding time). In general, the inter-arrival time is a function of
the colloid shape and radius, diffusion coefficient, and concentration
of colloids in the bulk. For simplicity, we consider equisized spherical
particles and the limit of highly diluted colloids. In this limit, the
diffusion coefficient affects only the trajectory of the colloids and
the inter-arrival time can be considered much larger than the binding
time. Since the binding is irreversible, we assume that the colloids
arrive one at a time towards the substrate and adhere instantaneously.
In addition, we consider chemical bonds between the patches, which are
highly directional and assumed irreversible within the timescale of
interest. In order to describe the colloid motion in solution, a
Brownian algorithm is considered, adapting ideas previously implemented
for Molecular Dynamics (see, e.g.  Ref.~\cite{Russo2009}). Collisions with 
the solvent are assumed Poisson processes, i.e., the
time between collisions is exponentially distributed. At each
particle-solvent collision, a new value for the particle velocity is
generated, drawn randomly from the Maxwell-Boltzmann distribution at the
thermostat temperature. The direction of the velocity is obtained from a
uniform distribution. By selecting the thermostat temperature and the
collision rate (Poisson process) we can then adjust the diffusion
coefficient. 
The substrate is considered attractive and, therefore, collisions with the substrate
always result in irreversible colloidal binding.

\begin{figure}[t]
  \begin{center}
  \includegraphics[width=\columnwidth]{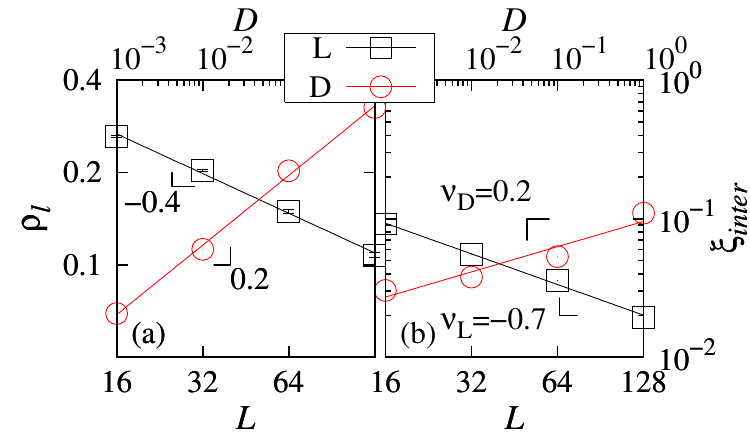} \\
  \end{center}
  \caption{
  (color online) (a) Liquid-film density for substrates with lateral size ranging from $16$ to $128$, and diffusion coefficient, $D$, between
$10^{-3}$ and $1$. (b) Dependence of the interfacial decay length $\xi_{\mathrm{inter}}$ on $L$ and $D$.
  \label{fig.figure4}}
\end{figure}

Our model shares important features with Diffusion Limited Deposition
(DLD), an extension of the famous DLA developed to account for the
growth of films on fibers and surfaces \cite{Meakin1998}. In DLD,
particles diffuse one after the other and adhere to the first
pre-adsorbed particle. For patchy colloids, since a link is established
only when there is an effective overlap between interaction ranges, the
colloid does not, necessarily, bind during the first collision. Besides,
somewhat artificially, in the continuum version of DLD the random
walkers' mean free path is considered of uniform length (typically
of the order of the diameter of the particle) and, consequently, the
diffusion coefficient only affects the time scale.  Here, we use a more
realistic description where the inter-collision time with the solvent,
and consequently the mean free path, is a function of the
thermostat temperature. As we will show below, an interesting dependence
on the diffusion coefficient emerges.

\section{Results}\label{sec.results}

\begin{figure}[t]
  \includegraphics[width=\columnwidth]{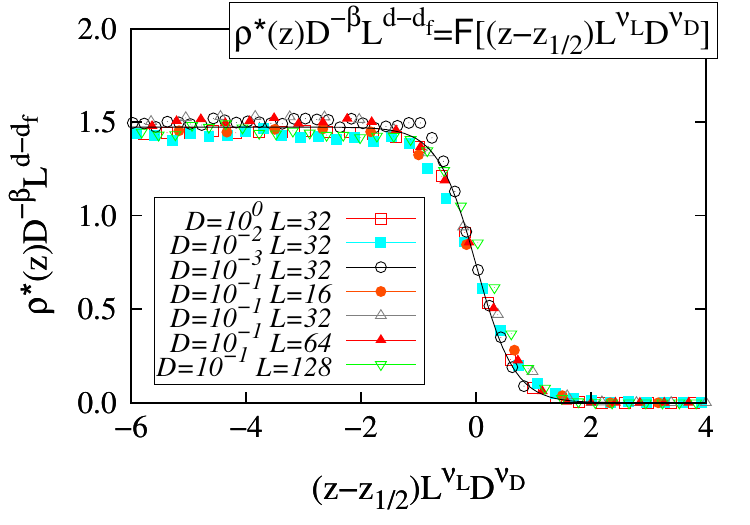}
  \caption{
(color online) Data collapse for the liquid film and interfacial region, 
including the dependence on the substrate lateral size $L$ and
the diffusion coefficient $D$. Results obtained after adsorbing $60$
particle layers and averaging over $500$ samples for $L=\{16,32\}$, $200$
samples for $L=64$, and $100$ samples for $L=128$.
  \label{fig.figure5}}
\end{figure}

We start with an empty planar square substrate, with lateral size $L$,
defined in units of the particle diameter, and assume periodic boundary
conditions in the horizontal plane. Iteratively, spherical colloids,
with three patches uniformly distributed on their surface, are released
one after the other and diffuse until they bind either to the substrate
or to a previously adsorbed colloid.  Figure~\ref{fig.figure2} shows a
snapshot of a typical network of connected three-patch colloids and the
density profile $\rho^*(z)$, defined as the number of particles per unit
volume, where $z$ is the distance to the substrate. One can
distinguish three different regimes: surface layer, liquid film, and
interfacial region. As we move away from the surface, the density
rapidly decreases (surface layer, $z<50$) until a saturation value
$\rho_l$, which is constant within the liquid film ($50<z<150$), and
vanishes in the interfacial region ($z>150$). In the following, we
discuss in detail each of these regimes.

 \textit{Surface layer.} As the colloid-substrate interaction is
isotropic, the patches of colloids directly adsorbed on the substrate
are oriented randomly. By contrast, the colloid-colloid (i.e.,
patch-patch) interaction is strongly anisotropic and the network chains
extend only along the direction of the patches. As the front of the
network propagates, its branches screen the inner layers and span along
the lateral direction, with a consequent decrease of the density. This
effect was also reported for DLD, where a power-law decay with $z$ is
observed, with multifractal scaling \cite{Meakin1998, Coniglio1990}.
However, here the conspiracy between the order promoted by the patches
and the disorder imposed by the first layer results in an exponential
decrease, at least for the lateral sizes that were investigated.

\begin{figure}[t]
  \begin{center}
  \includegraphics[width=\columnwidth]{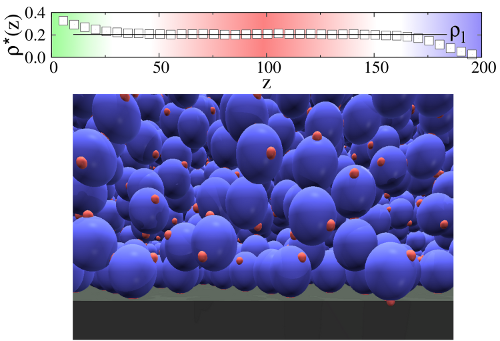} \\
  \end{center}
  \caption{
   (color online) \textit{Top:} Density profile of the colloidal network on a substrate with
anisotropic particle-substrate interaction. Results after the adsorption
of $40$ particle layers, averaged over $500$ independent realizations.
$\rho_l$ is the density of the liquid film. \textit{Bottom:} Snapshot of
the particles in the surface layer regime.
   \label{fig.figure6}}
\end{figure}

The plot in Fig.~\ref{fig.figure3} shows the dependence of the
first-layer density, $\rho^*(0)$, on the size of the substrate $L$ and
diffusion coefficient $D$. While there is no significant finite-size
effect, an increase of $\rho^*(0)$ with $D$ is observed (see snapshots
in the same figure). For the irreversible adhesion of colloids on
substrates, in the limit where particles only stick to the substrate and
not on top of other particles, extensive simulations have shown no
significant dependence on the diffusion coefficient \cite{Senger1991,
Senger1992}.  In that case, the structure of the film should resemble
that of Random Sequential Adsorption (RSA) \cite{Evans1993a,
Privman2000, Cadilhe2007}. Instead, the patch-patch interaction promotes
the formation of colloid chains hindering the access to the substrate.
Additionally, in the diffusion process, the typical colloid mean
free path increases with $D$. Since multiple collisions can occur
before irreversible binding, the larger the mean free path the higher the
probability that the colloids can squeeze into the fjords and,
eventually, arrive at the substrate. Consequently, the first-layer
density increases with $D$ towards the RSA limit (limit line in
Fig.~\ref{fig.figure3}). However, patchy particle systems will always
form networks and the RSA limit is never reached.

\begin{figure}[t]
  \begin{center}
  \includegraphics[width=\columnwidth]{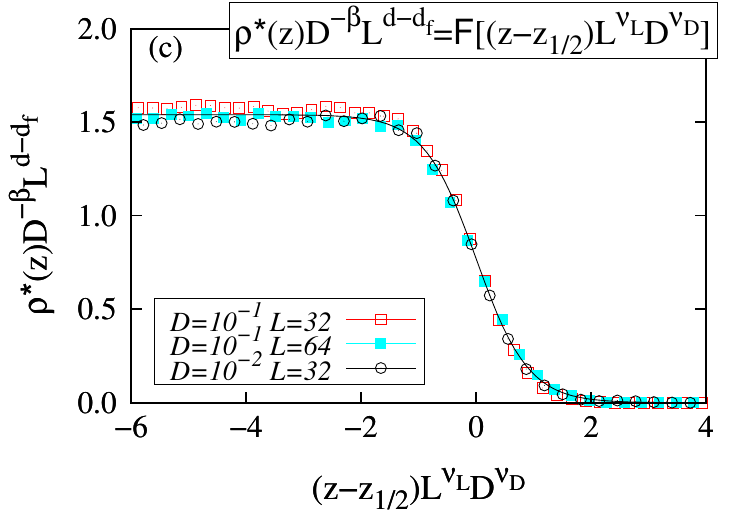} \\
  \end{center}
  \caption{
  (color online) Data collapse for the liquid film and interfacial region, where bonds are considered non-optimal. Results are obtained after adsorbing
$60$ particle layers and averaging over 500 samples for $L=32$ and 200
for $L=64$.
 \label{fig.figure7}}
\end{figure}

\textit{Liquid film.} As the lateral growth of the network proceeds,
the finite size of the substrate induces a saturation of the density at
$\rho=\rho_l$. For the range of values of $L$ and $D$ considered here,
the network is always a fractal with fractal dimension $d_f=2.58\pm0.04$,
calculated using the box counting algorithm; this value is compatible
with the one reported for DLA and DLD \cite{Meakin1998}.
Notwithstanding, $\rho_l$ depends on $L$ and $D$ (see
Fig.~\ref{fig.figure4}(a)). In both cases, a
power-law scaling is observed, namely, $\rho_l\sim L^{d_f-d}D^{\beta}$,
with $d_f-d=-0.4\pm0.1$ and $\beta=0.23\pm0.02$. While the exponent in
the size dependence results from the scale invariance of the network and
is straightforwardly connected with $d_f$, the dynamic exponent $\beta$
is new and there is no equivalent in previously studied models.

\begin{figure}[t]	
  \includegraphics[width=\columnwidth]{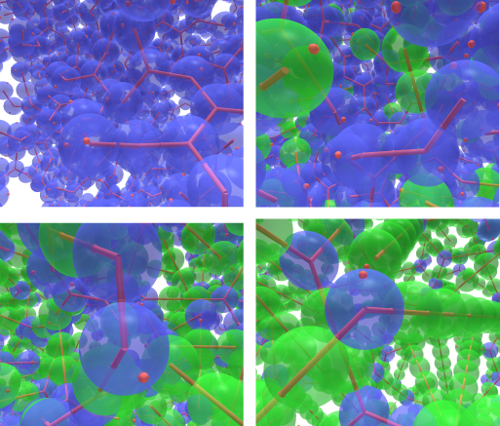}
  \caption{ 
(color online) Snapshots of a region in the liquid film for different
fractions of two-patch colloids, $r_D$. From left to right, top to
bottom: $0$,
$0.3$, $0.6$, and $0.8$. Snapshots obtained from the adsorption on a
substrate with a lateral size of $32$ particle diameters and $40$ particle
layers. Three-patch colloids are in blue (dark), two-patch colloids are in
green (light), the (red) spheres on the surface of the colloids represent 
the patches, and the (red) sticks are the connections between colloids.
  \label{fig.figure8}}
\end{figure}

\begin{figure}[t]
  \begin{center}
  \includegraphics[width=\columnwidth]{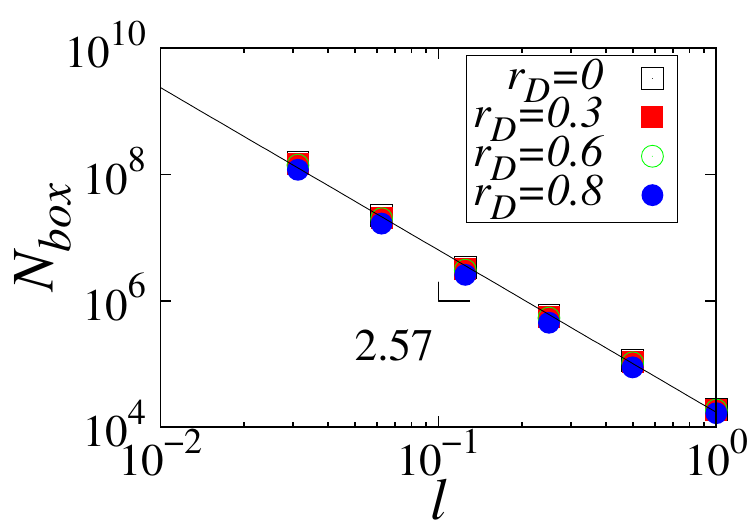} \\
  \end{center}
  \caption{
  (color online) Mass scaling as a function of the inverse box size in the box counting
algorithm, for networks, with different concentrations of two-patch colloids,
$r_D$. Results obtained with $32$ layers of the $60$ layers of adsorbed particles in the liquid film
on a substrate with $L=32$, averaged over $20$ samples. A fractal dimension $d_f=2.57\pm0.04$ was found.
  \label{fig.figure9}}
\end{figure}

\textit{Interfacial region.} The density vanishes at the interface
between the network and the solution, which corresponds to the active
front of the film. The position of the interface depends on the total
number of adsorbed particle layers $N$. We define $z_{1/2}$ as the
thickness of the film, corresponding to the height at which the density
is $\rho_l/2$ (see Fig.~\ref{fig.figure2}). Since the thickness of the
surface layer does not depend on $N$, $z_{1/2}$ asymptotically scales as
\mbox{$z_{1/2}\sim N/\rho_l=N L^{d-d_f}$}. The profile of the density in the
interfacial region scales as
\mbox{$\rho^*(z)=\rho_l\tanh\left[(z-z_{1/2})\xi_{\mathrm{inter}}\right]$},
as typically observed for models of diffusion limited growth in the
stationary regime (see, e.g., Ref.~\cite{Collins1985}).  The decay
exponent $\xi_{\mathrm{inter}}$ also scales with $L$ and $D$ (see 
Fig.~\ref{fig.figure4}(b)). In both cases, a power
law is observed, $\xi_{\mathrm{inter}}\sim L^{\nu_L}D^{\nu_D}$, with
$\nu_L=-0.73\pm0.04$ and $\nu_D=0.18\pm0.04$.

Based on the properties of the liquid film and interfacial region, we
propose full scaling of the density profile with $L$ and $D$, defined as
\mbox{$\rho^*(z)=D^\beta
L^{d_f-d}F\left[(z-z_{1/2})L^{\nu_L}D^{\nu_D}\right]$}, where
$F\left[x\right]$ is a scaling function described by the hyperbolic
tangent. Figure~\ref{fig.figure5} shows the data collapse for several
values of $D$ and $L$. This scaling allows the definition of the network
density for any size of the substrate and of the diffusion coefficient.

For simplicity, we have considered an isotropic particle-substrate
interaction. In general, some anisotropy is expected due to the patches.
Figure~\ref{fig.figure6} shows the density profile of the colloidal
network when this interaction is anisotropic. In particular, as
illustrated in the snapshot (bottom of the figure), we considered the case where
adsorbing colloids bind to the substrate only through the patches. The
parameters are those considered in Fig.~\ref{fig.figure2}. We show that the
qualitative picture discussed here is not affected by the details of the
interaction with the substrate and the quantitative results are
different only within the surface layer. 

We have also assumed optimal bonds such that binding is established along the direction of the
patches. It has been shown recently that non-optimal bonds may enrich the
diagram of self-organized states \cite{Romano2012}. In Fig.\ref{fig.figure7}, 
we also consider the case of non-optimal bonds, i.e., as in the optimal
case a landing patchy colloid only sticks to a previously adsorbed one
when their interaction ranges overlap, but the binding is established at
the contact point. In spite of changes in the liquid film density,
the same scaling and fractal dimensions are found.

\begin{figure}[t]	
  \includegraphics[width=\columnwidth]{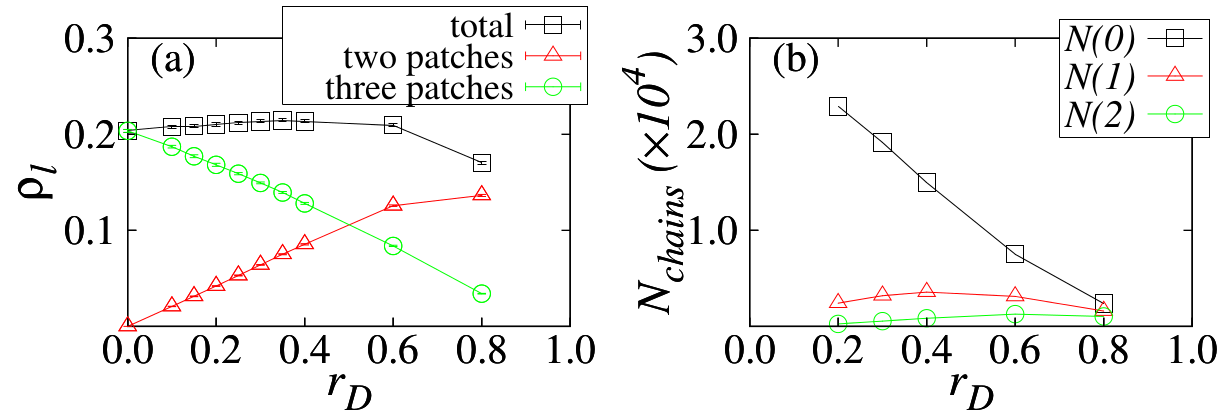}
  \caption{
(color online) Analysis of the dependence on the fraction of two-patch
colloids, $r_D$. (a) Liquid-film density (black squares)
and the contribution from two- (red triangles) and three-patch (green spheres) colloids. 
(b) Total number of chains of two-patch colloids between three-patch ones with sizes $0$, $1$, and $2$. 
The peak of the distribution of chains of unit size coincides with the peak of the film density. Results 
for $60$ particle layers adsorbed on a substrate with $L=32$ averaged over $500$ samples.
  \label{fig.figure10}}
\end{figure}

\textit{Adsorption of two- and three-patch colloids.} Finally we
consider the adsorption of a mixture of two- and three-patch colloids.
While the two-patch colloids (patches on the poles) favor long chains,
the three-patch particles promote branching. We define $r_D$ as the
fraction of adsorbed two-patch colloids and investigate how it affects
the network structure.  Figure~\ref{fig.figure8} shows snapshots of the
liquid film for different $r_D$. The larger $r_D$ the longer the chains
of two-patch colloids. In all cases, including $r_D=0$, the resulting
network is a fractal with the same fractal dimension (see
Fig.~\ref{fig.figure9}). However, the density of
the liquid film can be controlled with $r_D$ (see Fig.~\ref{fig.figure10}(a)). 

\begin{figure}[hp]
  \begin{center}
  \begin{tabular}{cc}
  \includegraphics[width=0.5\columnwidth]{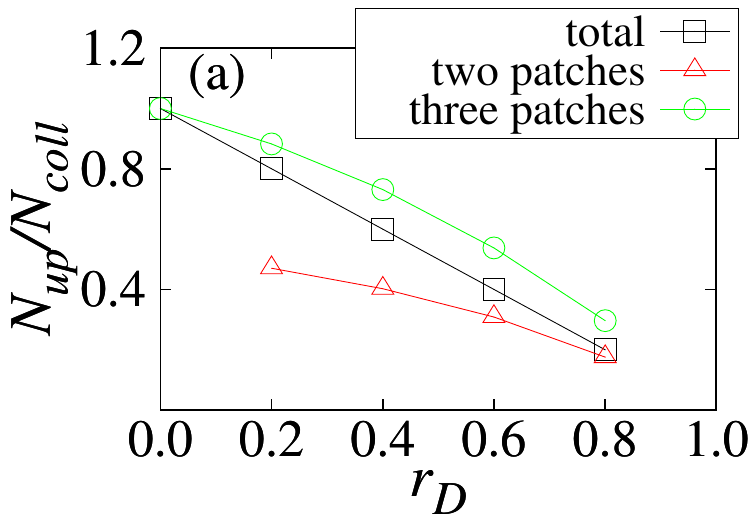} &
\includegraphics[width=0.5\columnwidth]{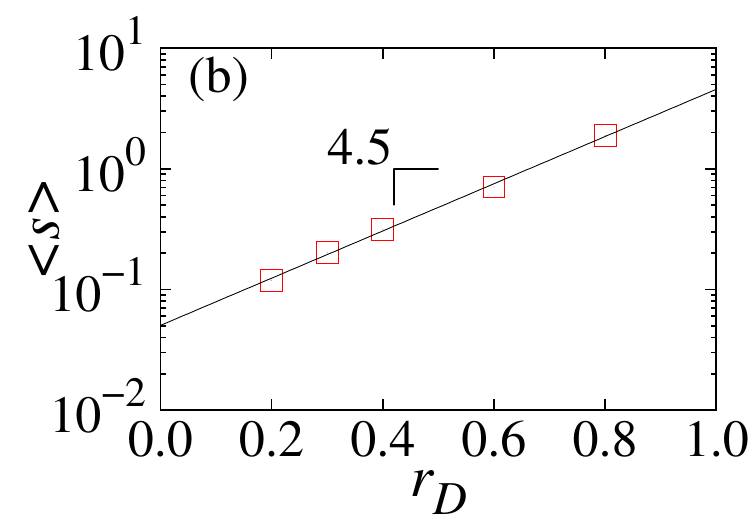} \\
  \end{tabular}
  \end{center}
  \caption{
  (color online) (a) Fraction of free (unconnected) patches on two-patch (red triangles) and three-patch (green circles) colloids in the liquid film. 
(b) Exponential increase of the average size of chains of two-patch colloids.
  \label{fig.figure11}}
\end{figure}

Somewhat surprisingly, a maximum is
observed in the density at a fraction of two-patch particles around
$0.35$, by contrast to equilibrium coexisting gels (i.e. optimal
networks at zero pressure) where a monotonic decrease is observed
\cite{Bianchi2006}. In fact, the film density remains above the
three-patch colloid limit ($r_D=0$) over a wide range of $r_D$ (up to
around $0.6$). As $r_D$ increases, long chains of two-patch colloids are
formed and the density is expected to decrease. However, in the absence
of relaxation, kinetically trapped structures are obtained and
geometrical constraints hinder the access of colloids to the
free-patches in the inner layers. The competition between the formation
of long chains and the maximization of accessible patches drives the
reported film density maximum. The plot in Fig.~\ref{fig.figure10}(b)
shows the number of chains of two-patch colloids of size $0$, $1$, and
$2$. With increasing $r_D$, the number of chains of size $0$
(corresponding to pairs of connected three-patch colloids) decreases and
a maximum in the number of chains of size $1$ is observed at the same
concentration as the maximum in the film density. In Fig.\ref{fig.figure11} we show that, increasing
$r_D$ reduces the number of free patches in the film and increases the
average size of the chains exponentially.

\section{Conclusions}\label{sec.conclusions} 

This work reveals that, in the presence of a
substrate and under nonequilibrium conditions, new self-organized
patterns are obtained which differ from the thermodynamic optimal
networks or equilibrium coexisting gels (at zero pressure). For mixtures
of three- and two-patch colloids, fractal networks of connected
particles are formed with a fractal dimension resilient over a wide
range of diffusion coefficients and concentration of two-patch colloids.
Yet, the obtained structures depend both on the diffusion coefficient
and the colloidal valence. These networks might be of relevance in the
fields of microfluidics and filtering. For example, the density
variation in the surface layer resembles the filtering mechanism
recently found in the human airways \cite{Button2012}. Here, we focused
on the nonequilibrium properties of the network, resulting from the
irreversible nature of the binding, which dominates at low temperatures.
However, in a significant long timescale or at higher temperatures, an
adsorbed colloid might detach and rebind to another patch or to the
substrate. In this case, the thermodynamic equilibrium structures might,
in principle, be reached.  As a follow up, the stability and aging of
the networks should be investigated as well as the kinetic pathways
towards the thermodynamic equilibrium structures.  Additionally,
techniques to stabilize these structures over extended periods of time
might also be a focus of future research. The model may be used to
investigate other features such as different particle-substrate
interactions \cite{Araujo2008,Wilner2012} and non-optimal bonds, where
fluctuations in the bonding direction are taken into account
\cite{Romano2012}.

\acknowledgments{ We acknowledge financial support from the
Portuguese Foundation for Science and Technology (FCT) under Contracts
nos. EXCL/FIS-NAN/0083/2012, PEst-OE/FIS/UI0618/2011, and
PTDC/FIS/098254/2008, and also helpful comments from Francisco de los
Santos. This work was also supported (NA) by grant number FP7-319968 
of the European Research Council}

  \bibliography{networkpatchy}

\begin{thebibliography}{43}%
\makeatletter
\providecommand \@ifxundefined [1]{%
 \@ifx{#1\undefined}
}%
\providecommand \@ifnum [1]{%
 \ifnum #1\expandafter \@firstoftwo
 \else \expandafter \@secondoftwo
 \fi
}%
\providecommand \@ifx [1]{%
 \ifx #1\expandafter \@firstoftwo
 \else \expandafter \@secondoftwo
 \fi
}%
\providecommand \natexlab [1]{#1}%
\providecommand \enquote  [1]{``#1''}%
\providecommand \bibnamefont  [1]{#1}%
\providecommand \bibfnamefont [1]{#1}%
\providecommand \citenamefont [1]{#1}%
\providecommand \href@noop [0]{\@secondoftwo}%
\providecommand \href [0]{\begingroup \@sanitize@url \@href}%
\providecommand \@href[1]{\@@startlink{#1}\@@href}%
\providecommand \@@href[1]{\endgroup#1\@@endlink}%
\providecommand \@sanitize@url [0]{\catcode `\\12\catcode `\$12\catcode
  `\&12\catcode `\#12\catcode `\^12\catcode `\_12\catcode `\%12\relax}%
\providecommand \@@startlink[1]{}%
\providecommand \@@endlink[0]{}%
\providecommand \url  [0]{\begingroup\@sanitize@url \@url }%
\providecommand \@url [1]{\endgroup\@href {#1}{\urlprefix }}%
\providecommand \urlprefix  [0]{URL }%
\providecommand \Eprint [0]{\href }%
\providecommand \doibase [0]{http://dx.doi.org/}%
\providecommand \selectlanguage [0]{\@gobble}%
\providecommand \bibinfo  [0]{\@secondoftwo}%
\providecommand \bibfield  [0]{\@secondoftwo}%
\providecommand \translation [1]{[#1]}%
\providecommand \BibitemOpen [0]{}%
\providecommand \bibitemStop [0]{}%
\providecommand \bibitemNoStop [0]{.\EOS\space}%
\providecommand \EOS [0]{\spacefactor3000\relax}%
\providecommand \BibitemShut  [1]{\csname bibitem#1\endcsname}%
\let\auto@bib@innerbib\@empty
\bibitem [{\citenamefont {Shum}\ \emph {et~al.}(2010)\citenamefont {Shum},
  \citenamefont {Abate}, \citenamefont {Lee}, \citenamefont {Studart},
  \citenamefont {Wang}, \citenamefont {Chen}, \citenamefont {Thiele},
  \citenamefont {Shah}, \citenamefont {Krummel},\ and\ \citenamefont
  {Weitz}}]{Shum2010}%
  \BibitemOpen
  \bibfield  {author} {\bibinfo {author} {\bibfnamefont {H.~C.}\ \bibnamefont
  {Shum}}, \bibinfo {author} {\bibfnamefont {A.~R.}\ \bibnamefont {Abate}},
  \bibinfo {author} {\bibfnamefont {D.}~\bibnamefont {Lee}}, \bibinfo {author}
  {\bibfnamefont {A.~R.}\ \bibnamefont {Studart}}, \bibinfo {author}
  {\bibfnamefont {B.}~\bibnamefont {Wang}}, \bibinfo {author} {\bibfnamefont
  {C.-H.}\ \bibnamefont {Chen}}, \bibinfo {author} {\bibfnamefont
  {J.}~\bibnamefont {Thiele}}, \bibinfo {author} {\bibfnamefont {R.~K.}\
  \bibnamefont {Shah}}, \bibinfo {author} {\bibfnamefont {A.}~\bibnamefont
  {Krummel}}, \ and\ \bibinfo {author} {\bibfnamefont {D.~A.}\ \bibnamefont
  {Weitz}},\ }\href@noop {} {\bibfield  {journal} {\bibinfo  {journal}
  {Macromol. Rapid Commun.}\ }\textbf {\bibinfo {volume} {31}},\ \bibinfo
  {pages} {108} (\bibinfo {year} {2010})}\BibitemShut {NoStop}%
\bibitem [{\citenamefont {Duguet}\ \emph {et~al.}(2011)\citenamefont {Duguet},
  \citenamefont {D\'{e}sert}, \citenamefont {Perro},\ and\ \citenamefont
  {Ravaine}}]{Duguet2011}%
  \BibitemOpen
  \bibfield  {author} {\bibinfo {author} {\bibfnamefont {E.}~\bibnamefont
  {Duguet}}, \bibinfo {author} {\bibfnamefont {A.}~\bibnamefont {D\'{e}sert}},
  \bibinfo {author} {\bibfnamefont {A.}~\bibnamefont {Perro}}, \ and\ \bibinfo
  {author} {\bibfnamefont {S.}~\bibnamefont {Ravaine}},\ }\href@noop {}
  {\bibfield  {journal} {\bibinfo  {journal} {Chem. Soc. Rev.}\ }\textbf
  {\bibinfo {volume} {40}},\ \bibinfo {pages} {941} (\bibinfo {year}
  {2011})}\BibitemShut {NoStop}%
\bibitem [{\citenamefont {Sacanna}\ and\ \citenamefont
  {Pine}(2011)}]{Sacanna2011}%
  \BibitemOpen
  \bibfield  {author} {\bibinfo {author} {\bibfnamefont {S.}~\bibnamefont
  {Sacanna}}\ and\ \bibinfo {author} {\bibfnamefont {D.~J.}\ \bibnamefont
  {Pine}},\ }\href@noop {} {\bibfield  {journal} {\bibinfo  {journal} {Curr.
  Opin. Coll. Interf. Sci.}\ }\textbf {\bibinfo {volume} {16}},\ \bibinfo
  {pages} {96} (\bibinfo {year} {2011})}\BibitemShut {NoStop}%
\bibitem [{\citenamefont {He}\ and\ \citenamefont
  {Kretzschmar}(2012)}]{He2012}%
  \BibitemOpen
  \bibfield  {author} {\bibinfo {author} {\bibfnamefont {Z.}~\bibnamefont
  {He}}\ and\ \bibinfo {author} {\bibfnamefont {I.}~\bibnamefont
  {Kretzschmar}},\ }\href@noop {} {\bibfield  {journal} {\bibinfo  {journal}
  {Langmuir}\ }\textbf {\bibinfo {volume} {28}},\ \bibinfo {pages} {9915}
  (\bibinfo {year} {2012})}\BibitemShut {NoStop}%
\bibitem [{\citenamefont {Hu}\ \emph {et~al.}(2012)\citenamefont {Hu},
  \citenamefont {Zhou}, \citenamefont {Sun}, \citenamefont {Fang},\ and\
  \citenamefont {Wu}}]{Hu2012}%
  \BibitemOpen
  \bibfield  {author} {\bibinfo {author} {\bibfnamefont {J.}~\bibnamefont
  {Hu}}, \bibinfo {author} {\bibfnamefont {S.}~\bibnamefont {Zhou}}, \bibinfo
  {author} {\bibfnamefont {Y.}~\bibnamefont {Sun}}, \bibinfo {author}
  {\bibfnamefont {X.}~\bibnamefont {Fang}}, \ and\ \bibinfo {author}
  {\bibfnamefont {L.}~\bibnamefont {Wu}},\ }\href@noop {} {\bibfield  {journal}
  {\bibinfo  {journal} {Chem. Soc. Rev.}\ }\textbf {\bibinfo {volume} {41}},\
  \bibinfo {pages} {4356} (\bibinfo {year} {2012})}\BibitemShut {NoStop}%
\bibitem [{\citenamefont {Wilner}\ and\ \citenamefont
  {Willner}(2012)}]{Wilner2012}%
  \BibitemOpen
  \bibfield  {author} {\bibinfo {author} {\bibfnamefont {O.~I.}\ \bibnamefont
  {Wilner}}\ and\ \bibinfo {author} {\bibfnamefont {I.}~\bibnamefont
  {Willner}},\ }\href@noop {} {\bibfield  {journal} {\bibinfo  {journal} {Chem.
  Rev.}\ }\textbf {\bibinfo {volume} {112}},\ \bibinfo {pages} {2528} (\bibinfo
  {year} {2012})}\BibitemShut {NoStop}%
\bibitem [{\citenamefont {Pawar}\ and\ \citenamefont
  {Kretzschmar}(2010)}]{Pawar2010}%
  \BibitemOpen
  \bibfield  {author} {\bibinfo {author} {\bibfnamefont {A.~B.}\ \bibnamefont
  {Pawar}}\ and\ \bibinfo {author} {\bibfnamefont {I.}~\bibnamefont
  {Kretzschmar}},\ }\href@noop {} {\bibfield  {journal} {\bibinfo  {journal}
  {Macromol. Rapid Commun.}\ }\textbf {\bibinfo {volume} {31}},\ \bibinfo
  {pages} {150} (\bibinfo {year} {2010})}\BibitemShut {NoStop}%
\bibitem [{\citenamefont {Kretzschmar}\ and\ \citenamefont
  {Song}(2011)}]{Kretzschmar2011}%
  \BibitemOpen
  \bibfield  {author} {\bibinfo {author} {\bibfnamefont {I.}~\bibnamefont
  {Kretzschmar}}\ and\ \bibinfo {author} {\bibfnamefont {J.~H.~K.}\
  \bibnamefont {Song}},\ }\href@noop {} {\bibfield  {journal} {\bibinfo
  {journal} {Curr. Opin. Coll. Interf. Sci.}\ }\textbf {\bibinfo {volume}
  {16}},\ \bibinfo {pages} {84} (\bibinfo {year} {2011})}\BibitemShut {NoStop}%
\bibitem [{\citenamefont {Bianchi}\ \emph {et~al.}(2011)\citenamefont
  {Bianchi}, \citenamefont {Blaak},\ and\ \citenamefont {Likos}}]{Bianchi2011}%
  \BibitemOpen
  \bibfield  {author} {\bibinfo {author} {\bibfnamefont {E.}~\bibnamefont
  {Bianchi}}, \bibinfo {author} {\bibfnamefont {R.}~\bibnamefont {Blaak}}, \
  and\ \bibinfo {author} {\bibfnamefont {C.~N.}\ \bibnamefont {Likos}},\
  }\href@noop {} {\bibfield  {journal} {\bibinfo  {journal} {Phys. Chem. Chem.
  Phys.}\ }\textbf {\bibinfo {volume} {13}},\ \bibinfo {pages} {6397} (\bibinfo
  {year} {2011})}\BibitemShut {NoStop}%
\bibitem [{\citenamefont {Russo}\ \emph {et~al.}(2009)\citenamefont {Russo},
  \citenamefont {Tartaglia},\ and\ \citenamefont {Sciortino}}]{Russo2009}%
  \BibitemOpen
  \bibfield  {author} {\bibinfo {author} {\bibfnamefont {J.}~\bibnamefont
  {Russo}}, \bibinfo {author} {\bibfnamefont {P.}~\bibnamefont {Tartaglia}}, \
  and\ \bibinfo {author} {\bibfnamefont {F.}~\bibnamefont {Sciortino}},\
  }\href@noop {} {\bibfield  {journal} {\bibinfo  {journal} {J. Chem. Phys.}\
  }\textbf {\bibinfo {volume} {131}},\ \bibinfo {pages} {014504} (\bibinfo
  {year} {2009})}\BibitemShut {NoStop}%
\bibitem [{\citenamefont {Sciortino}\ and\ \citenamefont
  {Zaccarelli}(2011)}]{Sciortino2011}%
  \BibitemOpen
  \bibfield  {author} {\bibinfo {author} {\bibfnamefont {F.}~\bibnamefont
  {Sciortino}}\ and\ \bibinfo {author} {\bibfnamefont {E.}~\bibnamefont
  {Zaccarelli}},\ }\href@noop {} {\bibfield  {journal} {\bibinfo  {journal}
  {Curr. Opin. Solid State Mater. Sci.}\ }\textbf {\bibinfo {volume} {15}},\
  \bibinfo {pages} {246} (\bibinfo {year} {2011})}\BibitemShut {NoStop}%
\bibitem [{\citenamefont {Sciortino}\ \emph {et~al.}(2007)\citenamefont
  {Sciortino}, \citenamefont {Bianchi}, \citenamefont {Douglas},\ and\
  \citenamefont {Tartaglia}}]{Sciortino2007}%
  \BibitemOpen
  \bibfield  {author} {\bibinfo {author} {\bibfnamefont {F.}~\bibnamefont
  {Sciortino}}, \bibinfo {author} {\bibfnamefont {E.}~\bibnamefont {Bianchi}},
  \bibinfo {author} {\bibfnamefont {J.~F.}\ \bibnamefont {Douglas}}, \ and\
  \bibinfo {author} {\bibfnamefont {P.}~\bibnamefont {Tartaglia}},\ }\href@noop
  {} {\bibfield  {journal} {\bibinfo  {journal} {J. Chem. Phys.}\ }\textbf
  {\bibinfo {volume} {126}},\ \bibinfo {pages} {194903} (\bibinfo {year}
  {2007})}\BibitemShut {NoStop}%
\bibitem [{\citenamefont {Zhang}\ and\ \citenamefont
  {Glotzer}(2004)}]{Glotzer2004}%
  \BibitemOpen
  \bibfield  {author} {\bibinfo {author} {\bibfnamefont {Z.}~\bibnamefont
  {Zhang}}\ and\ \bibinfo {author} {\bibfnamefont {S.~C.}\ \bibnamefont
  {Glotzer}},\ }\href@noop {} {\bibfield  {journal} {\bibinfo  {journal} {Nano
  Lett.}\ }\textbf {\bibinfo {volume} {4}},\ \bibinfo {pages} {1407} (\bibinfo
  {year} {2004})}\BibitemShut {NoStop}%
\bibitem [{\citenamefont {Doye}\ \emph {et~al.}(2007)\citenamefont {Doye},
  \citenamefont {Louis}, \citenamefont {Lin}, \citenamefont {Allen},
  \citenamefont {Noya}, \citenamefont {Wilber}, \citenamefont {Kok},\ and\
  \citenamefont {Lyus}}]{Doye2007}%
  \BibitemOpen
  \bibfield  {author} {\bibinfo {author} {\bibfnamefont {J.~P.~K.}\
  \bibnamefont {Doye}}, \bibinfo {author} {\bibfnamefont {A.~A.}\ \bibnamefont
  {Louis}}, \bibinfo {author} {\bibfnamefont {I.-C.}\ \bibnamefont {Lin}},
  \bibinfo {author} {\bibfnamefont {L.~R.}\ \bibnamefont {Allen}}, \bibinfo
  {author} {\bibfnamefont {E.~G.}\ \bibnamefont {Noya}}, \bibinfo {author}
  {\bibfnamefont {A.~W.}\ \bibnamefont {Wilber}}, \bibinfo {author}
  {\bibfnamefont {H.~C.}\ \bibnamefont {Kok}}, \ and\ \bibinfo {author}
  {\bibfnamefont {R.}~\bibnamefont {Lyus}},\ }\href@noop {} {\bibfield
  {journal} {\bibinfo  {journal} {Phys. Chem. Chem. Phys.}\ }\textbf {\bibinfo
  {volume} {9}},\ \bibinfo {pages} {2197} (\bibinfo {year} {2007})}\BibitemShut
  {NoStop}%
\bibitem [{\citenamefont {Sciortino}\ \emph {et~al.}(2010)\citenamefont
  {Sciortino}, \citenamefont {Giacometti},\ and\ \citenamefont
  {Pastore}}]{Sciortino2010}%
  \BibitemOpen
  \bibfield  {author} {\bibinfo {author} {\bibfnamefont {F.}~\bibnamefont
  {Sciortino}}, \bibinfo {author} {\bibfnamefont {A.}~\bibnamefont
  {Giacometti}}, \ and\ \bibinfo {author} {\bibfnamefont {G.}~\bibnamefont
  {Pastore}},\ }\href@noop {} {\bibfield  {journal} {\bibinfo  {journal} {Phys.
  Chem. Chem. Phys.}\ }\textbf {\bibinfo {volume} {12}},\ \bibinfo {pages}
  {11869} (\bibinfo {year} {2010})}\BibitemShut {NoStop}%
\bibitem [{\citenamefont {Glotzer}\ and\ \citenamefont
  {Anderson}(2010)}]{Glotzer2010}%
  \BibitemOpen
  \bibfield  {author} {\bibinfo {author} {\bibfnamefont {S.~C.}\ \bibnamefont
  {Glotzer}}\ and\ \bibinfo {author} {\bibfnamefont {J.~A.}\ \bibnamefont
  {Anderson}},\ }\href@noop {} {\bibfield  {journal} {\bibinfo  {journal}
  {Nature Mater.}\ }\textbf {\bibinfo {volume} {9}},\ \bibinfo {pages} {885}
  (\bibinfo {year} {2010})}\BibitemShut {NoStop}%
\bibitem [{\citenamefont {Rosenthal}\ and\ \citenamefont
  {Klapp}(2011)}]{Rosenthal2011}%
  \BibitemOpen
  \bibfield  {author} {\bibinfo {author} {\bibfnamefont {G.}~\bibnamefont
  {Rosenthal}}\ and\ \bibinfo {author} {\bibfnamefont {S.~H.~L.}\ \bibnamefont
  {Klapp}},\ }\href@noop {} {\bibfield  {journal} {\bibinfo  {journal} {J.
  Chem. Phys.}\ }\textbf {\bibinfo {volume} {134}},\ \bibinfo {pages} {154707}
  (\bibinfo {year} {2011})}\BibitemShut {NoStop}%
\bibitem [{\citenamefont {Ruzicka}\ \emph {et~al.}(2010)\citenamefont
  {Ruzicka}, \citenamefont {Zaccarelli}, \citenamefont {Zulian}, \citenamefont
  {Angelini}, \citenamefont {Sztucki}, \citenamefont {Moussa\"{\i}d},
  \citenamefont {Narayanan},\ and\ \citenamefont {Sciortino}}]{Ruzicka2011}%
  \BibitemOpen
  \bibfield  {author} {\bibinfo {author} {\bibfnamefont {B.}~\bibnamefont
  {Ruzicka}}, \bibinfo {author} {\bibfnamefont {E.}~\bibnamefont {Zaccarelli}},
  \bibinfo {author} {\bibfnamefont {L.}~\bibnamefont {Zulian}}, \bibinfo
  {author} {\bibfnamefont {R.}~\bibnamefont {Angelini}}, \bibinfo {author}
  {\bibfnamefont {M.}~\bibnamefont {Sztucki}}, \bibinfo {author} {\bibfnamefont
  {A.}~\bibnamefont {Moussa\"{\i}d}}, \bibinfo {author} {\bibfnamefont
  {T.}~\bibnamefont {Narayanan}}, \ and\ \bibinfo {author} {\bibfnamefont
  {F.}~\bibnamefont {Sciortino}},\ }\href@noop {} {\bibfield  {journal}
  {\bibinfo  {journal} {Nature Mater.}\ }\textbf {\bibinfo {volume} {10}},\
  \bibinfo {pages} {56} (\bibinfo {year} {2010})}\BibitemShut {NoStop}%
\bibitem [{\citenamefont {Gnan}\ \emph {et~al.}(2012)\citenamefont {Gnan},
  \citenamefont {{de Las Heras}}, \citenamefont {Tavares}, \citenamefont {{Telo
  da Gama}},\ and\ \citenamefont {Sciortino}}]{Gnan2012}%
  \BibitemOpen
  \bibfield  {author} {\bibinfo {author} {\bibfnamefont {N.}~\bibnamefont
  {Gnan}}, \bibinfo {author} {\bibfnamefont {D.}~\bibnamefont {{de Las
  Heras}}}, \bibinfo {author} {\bibfnamefont {J.~M.}\ \bibnamefont {Tavares}},
  \bibinfo {author} {\bibfnamefont {M.~M.}\ \bibnamefont {{Telo da Gama}}}, \
  and\ \bibinfo {author} {\bibfnamefont {F.}~\bibnamefont {Sciortino}},\
  }\href@noop {} {\bibfield  {journal} {\bibinfo  {journal} {J. Chem. Phys.}\
  }\textbf {\bibinfo {volume} {137}},\ \bibinfo {pages} {084704} (\bibinfo
  {year} {2012})}\BibitemShut {NoStop}%
\bibitem [{\citenamefont {Bernardino}\ and\ \citenamefont {{Telo da
  Gama}}(2012)}]{Bernardino2012}%
  \BibitemOpen
  \bibfield  {author} {\bibinfo {author} {\bibfnamefont {N.~R.}\ \bibnamefont
  {Bernardino}}\ and\ \bibinfo {author} {\bibfnamefont {M.~M.}\ \bibnamefont
  {{Telo da Gama}}},\ }\href@noop {} {\bibfield  {journal} {\bibinfo  {journal}
  {Phys. Rev. Lett.}\ }\textbf {\bibinfo {volume} {109}},\ \bibinfo {pages}
  {116103} (\bibinfo {year} {2012})}\BibitemShut {NoStop}%
\bibitem [{\citenamefont {Cadilhe}\ \emph {et~al.}(2007)\citenamefont
  {Cadilhe}, \citenamefont {Ara\'{u}jo},\ and\ \citenamefont
  {Privman}}]{Cadilhe2007}%
  \BibitemOpen
  \bibfield  {author} {\bibinfo {author} {\bibfnamefont {A.}~\bibnamefont
  {Cadilhe}}, \bibinfo {author} {\bibfnamefont {N.~A.~M.}\ \bibnamefont
  {Ara\'{u}jo}}, \ and\ \bibinfo {author} {\bibfnamefont {V.}~\bibnamefont
  {Privman}},\ }\href@noop {} {\bibfield  {journal} {\bibinfo  {journal} {J.
  Phys.: Condens. Matter}\ }\textbf {\bibinfo {volume} {19}},\ \bibinfo {pages}
  {065124} (\bibinfo {year} {2007})}\BibitemShut {NoStop}%
\bibitem [{\citenamefont {Ara\'{u}jo}\ \emph {et~al.}(2008)\citenamefont
  {Ara\'{u}jo}, \citenamefont {Cadilhe},\ and\ \citenamefont
  {Privman}}]{Araujo2008}%
  \BibitemOpen
  \bibfield  {author} {\bibinfo {author} {\bibfnamefont {N.~A.~M.}\
  \bibnamefont {Ara\'{u}jo}}, \bibinfo {author} {\bibfnamefont
  {A.}~\bibnamefont {Cadilhe}}, \ and\ \bibinfo {author} {\bibfnamefont
  {V.}~\bibnamefont {Privman}},\ }\href@noop {} {\bibfield  {journal} {\bibinfo
   {journal} {Phys. Rev. E}\ }\textbf {\bibinfo {volume} {77}},\ \bibinfo
  {pages} {031603} (\bibinfo {year} {2008})}\BibitemShut {NoStop}%
\bibitem [{\citenamefont {Einstein}\ and\ \citenamefont
  {Stasevich}(2010)}]{Einstein2010}%
  \BibitemOpen
  \bibfield  {author} {\bibinfo {author} {\bibfnamefont {T.~L.}\ \bibnamefont
  {Einstein}}\ and\ \bibinfo {author} {\bibfnamefont {T.~J.}\ \bibnamefont
  {Stasevich}},\ }\href@noop {} {\bibfield  {journal} {\bibinfo  {journal}
  {Science}\ }\textbf {\bibinfo {volume} {327}},\ \bibinfo {pages} {423}
  (\bibinfo {year} {2010})}\BibitemShut {NoStop}%
\bibitem [{\citenamefont {Meakin}(1998)}]{Meakin1998}%
  \BibitemOpen
  \bibfield  {author} {\bibinfo {author} {\bibfnamefont {P.}~\bibnamefont
  {Meakin}},\ }\href@noop {} {\emph {\bibinfo {title} {{Fractals, scaling and
  growth far from equilibrium}}}}\ (\bibinfo  {publisher} {Cambridge Univ Pr},\
  \bibinfo {address} {Cambridge},\ \bibinfo {year} {1998})\BibitemShut
  {NoStop}%
\bibitem [{\citenamefont {{Witten Jr.}}\ and\ \citenamefont
  {Sander}(1981)}]{Witten1981}%
  \BibitemOpen
  \bibfield  {author} {\bibinfo {author} {\bibfnamefont {T.~A.}\ \bibnamefont
  {{Witten Jr.}}}\ and\ \bibinfo {author} {\bibfnamefont {L.~M.}\ \bibnamefont
  {Sander}},\ }\href@noop {} {\bibfield  {journal} {\bibinfo  {journal} {Phys.
  Rev. Lett.}\ }\textbf {\bibinfo {volume} {47}},\ \bibinfo {pages} {1400}
  (\bibinfo {year} {1981})}\BibitemShut {NoStop}%
\bibitem [{\citenamefont {Bianchi}\ \emph {et~al.}(2007)\citenamefont
  {Bianchi}, \citenamefont {Tartaglia}, \citenamefont {{La Nave}},\ and\
  \citenamefont {Sciortino}}]{Bianchi2007}%
  \BibitemOpen
  \bibfield  {author} {\bibinfo {author} {\bibfnamefont {E.}~\bibnamefont
  {Bianchi}}, \bibinfo {author} {\bibfnamefont {P.}~\bibnamefont {Tartaglia}},
  \bibinfo {author} {\bibfnamefont {E.}~\bibnamefont {{La Nave}}}, \ and\
  \bibinfo {author} {\bibfnamefont {F.}~\bibnamefont {Sciortino}},\ }\href@noop
  {} {\bibfield  {journal} {\bibinfo  {journal} {J. Phys. Chem. B}\ }\textbf
  {\bibinfo {volume} {111}},\ \bibinfo {pages} {11765} (\bibinfo {year}
  {2007})}\BibitemShut {NoStop}%
\bibitem [{\citenamefont {Bianchi}\ \emph {et~al.}(2008)\citenamefont
  {Bianchi}, \citenamefont {Tartaglia}, \citenamefont {Zaccarelli},\ and\
  \citenamefont {Sciortino}}]{Bianchi2008}%
  \BibitemOpen
  \bibfield  {author} {\bibinfo {author} {\bibfnamefont {E.}~\bibnamefont
  {Bianchi}}, \bibinfo {author} {\bibfnamefont {P.}~\bibnamefont {Tartaglia}},
  \bibinfo {author} {\bibfnamefont {E.}~\bibnamefont {Zaccarelli}}, \ and\
  \bibinfo {author} {\bibfnamefont {F.}~\bibnamefont {Sciortino}},\ }\href@noop
  {} {\bibfield  {journal} {\bibinfo  {journal} {J. Chem. Phys.}\ }\textbf
  {\bibinfo {volume} {128}},\ \bibinfo {pages} {144504} (\bibinfo {year}
  {2008})}\BibitemShut {NoStop}%
\bibitem [{\citenamefont {Bianchi}\ \emph {et~al.}(2006)\citenamefont
  {Bianchi}, \citenamefont {Largo}, \citenamefont {Tartaglia}, \citenamefont
  {Zaccarelli},\ and\ \citenamefont {Sciortino}}]{Bianchi2006}%
  \BibitemOpen
  \bibfield  {author} {\bibinfo {author} {\bibfnamefont {E.}~\bibnamefont
  {Bianchi}}, \bibinfo {author} {\bibfnamefont {J.}~\bibnamefont {Largo}},
  \bibinfo {author} {\bibfnamefont {P.}~\bibnamefont {Tartaglia}}, \bibinfo
  {author} {\bibfnamefont {E.}~\bibnamefont {Zaccarelli}}, \ and\ \bibinfo
  {author} {\bibfnamefont {F.}~\bibnamefont {Sciortino}},\ }\href@noop {}
  {\bibfield  {journal} {\bibinfo  {journal} {Phys. Rev. Lett.}\ }\textbf
  {\bibinfo {volume} {97}},\ \bibinfo {pages} {168301} (\bibinfo {year}
  {2006})}\BibitemShut {NoStop}%
\bibitem [{\citenamefont {Russo}\ and\ \citenamefont
  {Sciortino}(2010)}]{Russo2010}%
  \BibitemOpen
  \bibfield  {author} {\bibinfo {author} {\bibfnamefont {J.}~\bibnamefont
  {Russo}}\ and\ \bibinfo {author} {\bibfnamefont {F.}~\bibnamefont
  {Sciortino}},\ }\href@noop {} {\bibfield  {journal} {\bibinfo  {journal}
  {Phys. Rev. Lett.}\ }\textbf {\bibinfo {volume} {104}},\ \bibinfo {pages}
  {195701} (\bibinfo {year} {2010})}\BibitemShut {NoStop}%
\bibitem [{\citenamefont {Cho}\ \emph {et~al.}(2005)\citenamefont {Cho},
  \citenamefont {Yi}, \citenamefont {Lim}, \citenamefont {Kim}, \citenamefont
  {Manoharan}, \citenamefont {Pine},\ and\ \citenamefont {Yang}}]{Cho2005}%
  \BibitemOpen
  \bibfield  {author} {\bibinfo {author} {\bibfnamefont {Y.-S.}\ \bibnamefont
  {Cho}}, \bibinfo {author} {\bibfnamefont {G.-R.}\ \bibnamefont {Yi}},
  \bibinfo {author} {\bibfnamefont {J.-M.}\ \bibnamefont {Lim}}, \bibinfo
  {author} {\bibfnamefont {S.-H.}\ \bibnamefont {Kim}}, \bibinfo {author}
  {\bibfnamefont {V.-N.}\ \bibnamefont {Manoharan}}, \bibinfo {author}
  {\bibfnamefont {D.~J.}\ \bibnamefont {Pine}}, \ and\ \bibinfo {author}
  {\bibfnamefont {S.-M.}\ \bibnamefont {Yang}},\ }\href@noop {} {\bibfield
  {journal} {\bibinfo  {journal} {J. Am. Chem. Soc.}\ }\textbf {\bibinfo
  {volume} {127}},\ \bibinfo {pages} {15968} (\bibinfo {year}
  {2005})}\BibitemShut {NoStop}%
\bibitem [{\citenamefont {van Blaaderen}(2006)}]{Blaaderen2006}%
  \BibitemOpen
  \bibfield  {author} {\bibinfo {author} {\bibfnamefont {A.}~\bibnamefont {van
  Blaaderen}},\ }\href@noop {} {\bibfield  {journal} {\bibinfo  {journal}
  {Nature}\ }\textbf {\bibinfo {volume} {439}},\ \bibinfo {pages} {545}
  (\bibinfo {year} {2006})}\BibitemShut {NoStop}%
\bibitem [{\citenamefont {Kraft}\ \emph {et~al.}(2012)\citenamefont {Kraft},
  \citenamefont {Ni}, \citenamefont {Smallenburg}, \citenamefont {Hermes},
  \citenamefont {Yoon}, \citenamefont {Weitz}, \citenamefont {van Blaaderen},
  \citenamefont {Groenewold}, \citenamefont {Dijkstra},\ and\ \citenamefont
  {Kegel}}]{Kraft2012}%
  \BibitemOpen
  \bibfield  {author} {\bibinfo {author} {\bibfnamefont {D.~J.}\ \bibnamefont
  {Kraft}}, \bibinfo {author} {\bibfnamefont {R.}~\bibnamefont {Ni}}, \bibinfo
  {author} {\bibfnamefont {F.}~\bibnamefont {Smallenburg}}, \bibinfo {author}
  {\bibfnamefont {M.}~\bibnamefont {Hermes}}, \bibinfo {author} {\bibfnamefont
  {K.}~\bibnamefont {Yoon}}, \bibinfo {author} {\bibfnamefont {D.~A.}\
  \bibnamefont {Weitz}}, \bibinfo {author} {\bibfnamefont {A.}~\bibnamefont
  {van Blaaderen}}, \bibinfo {author} {\bibfnamefont {J.}~\bibnamefont
  {Groenewold}}, \bibinfo {author} {\bibfnamefont {M.}~\bibnamefont
  {Dijkstra}}, \ and\ \bibinfo {author} {\bibfnamefont {W.~K.}\ \bibnamefont
  {Kegel}},\ }\href@noop {} {\bibfield  {journal} {\bibinfo  {journal} {Proc.
  Nat. Acad. Sci.}\ }\textbf {\bibinfo {volume} {109}},\ \bibinfo {pages}
  {10787} (\bibinfo {year} {2012})}\BibitemShut {NoStop}%
\bibitem [{\citenamefont {Wang}\ \emph {et~al.}(2012)\citenamefont {Wang},
  \citenamefont {Breed}, \citenamefont {Manoharan}, \citenamefont {Feng},
  \citenamefont {Hollingsworth}, \citenamefont {Weck},\ and\ \citenamefont
  {Pine}}]{Wang2012}%
  \BibitemOpen
  \bibfield  {author} {\bibinfo {author} {\bibfnamefont {Y.}~\bibnamefont
  {Wang}}, \bibinfo {author} {\bibfnamefont {D.~R.}\ \bibnamefont {Breed}},
  \bibinfo {author} {\bibfnamefont {V.~N.}\ \bibnamefont {Manoharan}}, \bibinfo
  {author} {\bibfnamefont {L.}~\bibnamefont {Feng}}, \bibinfo {author}
  {\bibfnamefont {A.~D.}\ \bibnamefont {Hollingsworth}}, \bibinfo {author}
  {\bibfnamefont {M.}~\bibnamefont {Weck}}, \ and\ \bibinfo {author}
  {\bibfnamefont {D.~J.}\ \bibnamefont {Pine}},\ }\href@noop {} {\bibfield
  {journal} {\bibinfo  {journal} {Nature}\ }\textbf {\bibinfo {volume} {491}},\
  \bibinfo {pages} {51} (\bibinfo {year} {2012})}\BibitemShut {NoStop}%
\bibitem [{\citenamefont {Liu}\ \emph {et~al.}(2009)\citenamefont {Liu},
  \citenamefont {Kumar}, \citenamefont {Sciortino},\ and\ \citenamefont
  {Evans}}]{Liu2009}%
  \BibitemOpen
  \bibfield  {author} {\bibinfo {author} {\bibfnamefont {H.}~\bibnamefont
  {Liu}}, \bibinfo {author} {\bibfnamefont {S.~K.}\ \bibnamefont {Kumar}},
  \bibinfo {author} {\bibfnamefont {F.}~\bibnamefont {Sciortino}}, \ and\
  \bibinfo {author} {\bibfnamefont {G.~T.}\ \bibnamefont {Evans}},\ }\href@noop
  {} {\bibfield  {journal} {\bibinfo  {journal} {J. Chem. Phys.}\ }\textbf
  {\bibinfo {volume} {130}},\ \bibinfo {pages} {044902} (\bibinfo {year}
  {2009})}\BibitemShut {NoStop}%
\bibitem [{\citenamefont {Villar}\ \emph {et~al.}(2009)\citenamefont {Villar},
  \citenamefont {Wilber}, \citenamefont {Williamson}, \citenamefont {Thiara},
  \citenamefont {Doye}, \citenamefont {Louis}, \citenamefont {Jochum},
  \citenamefont {Lewis},\ and\ \citenamefont {Levy}}]{Villar2009}%
  \BibitemOpen
  \bibfield  {author} {\bibinfo {author} {\bibfnamefont {G.}~\bibnamefont
  {Villar}}, \bibinfo {author} {\bibfnamefont {A.~W.}\ \bibnamefont {Wilber}},
  \bibinfo {author} {\bibfnamefont {A.~J.}\ \bibnamefont {Williamson}},
  \bibinfo {author} {\bibfnamefont {P.}~\bibnamefont {Thiara}}, \bibinfo
  {author} {\bibfnamefont {J.~P.~K.}\ \bibnamefont {Doye}}, \bibinfo {author}
  {\bibfnamefont {A.~A.}\ \bibnamefont {Louis}}, \bibinfo {author}
  {\bibfnamefont {M.~N.}\ \bibnamefont {Jochum}}, \bibinfo {author}
  {\bibfnamefont {A.~C.~F.}\ \bibnamefont {Lewis}}, \ and\ \bibinfo {author}
  {\bibfnamefont {E.~D.}\ \bibnamefont {Levy}},\ }\href@noop {} {\bibfield
  {journal} {\bibinfo  {journal} {Phys. Rev. Lett.}\ }\textbf {\bibinfo
  {volume} {102}},\ \bibinfo {pages} {118106} (\bibinfo {year}
  {2009})}\BibitemShut {NoStop}%
\bibitem [{\citenamefont {Coniglio}\ and\ \citenamefont
  {Zannetti}(1990)}]{Coniglio1990}%
  \BibitemOpen
  \bibfield  {author} {\bibinfo {author} {\bibfnamefont {A.}~\bibnamefont
  {Coniglio}}\ and\ \bibinfo {author} {\bibfnamefont {M.}~\bibnamefont
  {Zannetti}},\ }\href@noop {} {\bibfield  {journal} {\bibinfo  {journal}
  {Physica A}\ }\textbf {\bibinfo {volume} {163}},\ \bibinfo {pages} {325}
  (\bibinfo {year} {1990})}\BibitemShut {NoStop}%
\bibitem [{\citenamefont {Senger}\ \emph {et~al.}(1991)\citenamefont {Senger},
  \citenamefont {Voegel}, \citenamefont {Schaaf}, \citenamefont {Johner},
  \citenamefont {Schmitt},\ and\ \citenamefont {Talbot}}]{Senger1991}%
  \BibitemOpen
  \bibfield  {author} {\bibinfo {author} {\bibfnamefont {B.}~\bibnamefont
  {Senger}}, \bibinfo {author} {\bibfnamefont {J.~C.}\ \bibnamefont {Voegel}},
  \bibinfo {author} {\bibfnamefont {P.}~\bibnamefont {Schaaf}}, \bibinfo
  {author} {\bibfnamefont {A.}~\bibnamefont {Johner}}, \bibinfo {author}
  {\bibfnamefont {A.}~\bibnamefont {Schmitt}}, \ and\ \bibinfo {author}
  {\bibfnamefont {J.}~\bibnamefont {Talbot}},\ }\href@noop {} {\bibfield
  {journal} {\bibinfo  {journal} {Phys. Rev. A}\ }\textbf {\bibinfo {volume}
  {44}},\ \bibinfo {pages} {6926} (\bibinfo {year} {1991})}\BibitemShut
  {NoStop}%
\bibitem [{\citenamefont {Senger}\ \emph {et~al.}(1992)\citenamefont {Senger},
  \citenamefont {Schaaf}, \citenamefont {Voegel}, \citenamefont {Johner},
  \citenamefont {Schmitt},\ and\ \citenamefont {Talbot}}]{Senger1992}%
  \BibitemOpen
  \bibfield  {author} {\bibinfo {author} {\bibfnamefont {B.}~\bibnamefont
  {Senger}}, \bibinfo {author} {\bibfnamefont {P.}~\bibnamefont {Schaaf}},
  \bibinfo {author} {\bibfnamefont {J.~C.}\ \bibnamefont {Voegel}}, \bibinfo
  {author} {\bibfnamefont {A.}~\bibnamefont {Johner}}, \bibinfo {author}
  {\bibfnamefont {A.}~\bibnamefont {Schmitt}}, \ and\ \bibinfo {author}
  {\bibfnamefont {J.}~\bibnamefont {Talbot}},\ }\href@noop {} {\bibfield
  {journal} {\bibinfo  {journal} {J. Chem. Phys.}\ }\textbf {\bibinfo {volume}
  {97}},\ \bibinfo {pages} {3813} (\bibinfo {year} {1992})}\BibitemShut
  {NoStop}%
\bibitem [{\citenamefont {Evans}(1993)}]{Evans1993a}%
  \BibitemOpen
  \bibfield  {author} {\bibinfo {author} {\bibfnamefont {J.~W.}\ \bibnamefont
  {Evans}},\ }\href@noop {} {\bibfield  {journal} {\bibinfo  {journal} {Rev.
  Mod. Phys.}\ }\textbf {\bibinfo {volume} {65}},\ \bibinfo {pages} {1281}
  (\bibinfo {year} {1993})}\BibitemShut {NoStop}%
\bibitem [{\citenamefont {Privman}(2000)}]{Privman2000}%
  \BibitemOpen
  \bibfield  {author} {\bibinfo {author} {\bibfnamefont {V.}~\bibnamefont
  {Privman}},\ }\href@noop {} {\bibfield  {journal} {\bibinfo  {journal} {J.
  Adhesion}\ }\textbf {\bibinfo {volume} {74}},\ \bibinfo {pages} {421}
  (\bibinfo {year} {2000})}\BibitemShut {NoStop}%
\bibitem [{\citenamefont {Collins}\ and\ \citenamefont
  {Levine}(1985)}]{Collins1985}%
  \BibitemOpen
  \bibfield  {author} {\bibinfo {author} {\bibfnamefont {J.~B.}\ \bibnamefont
  {Collins}}\ and\ \bibinfo {author} {\bibfnamefont {H.}~\bibnamefont
  {Levine}},\ }\href@noop {} {\bibfield  {journal} {\bibinfo  {journal} {Phys.
  Rev. B}\ }\textbf {\bibinfo {volume} {31}},\ \bibinfo {pages} {6119}
  (\bibinfo {year} {1985})}\BibitemShut {NoStop}%
\bibitem [{\citenamefont {Romano}\ and\ \citenamefont
  {Sciortino}(2012)}]{Romano2012}%
  \BibitemOpen
  \bibfield  {author} {\bibinfo {author} {\bibfnamefont {F.}~\bibnamefont
  {Romano}}\ and\ \bibinfo {author} {\bibfnamefont {F.}~\bibnamefont
  {Sciortino}},\ }\href@noop {} {\bibfield  {journal} {\bibinfo  {journal}
  {Nat. Comm.}\ }\textbf {\bibinfo {volume} {3}},\ \bibinfo {pages} {975}
  (\bibinfo {year} {2012})}\BibitemShut {NoStop}%
\bibitem [{\citenamefont {Button}\ \emph {et~al.}(2012)\citenamefont {Button},
  \citenamefont {Cai}, \citenamefont {Ehre}, \citenamefont {Kesimer},
  \citenamefont {Hill}, \citenamefont {Sheehan}, \citenamefont {Boucher},\ and\
  \citenamefont {Rubinstein}}]{Button2012}%
  \BibitemOpen
  \bibfield  {author} {\bibinfo {author} {\bibfnamefont {B.}~\bibnamefont
  {Button}}, \bibinfo {author} {\bibfnamefont {L.-H.}\ \bibnamefont {Cai}},
  \bibinfo {author} {\bibfnamefont {C.}~\bibnamefont {Ehre}}, \bibinfo {author}
  {\bibfnamefont {M.}~\bibnamefont {Kesimer}}, \bibinfo {author} {\bibfnamefont
  {D.~B.}\ \bibnamefont {Hill}}, \bibinfo {author} {\bibfnamefont {J.~K.}\
  \bibnamefont {Sheehan}}, \bibinfo {author} {\bibfnamefont {R.~C.}\
  \bibnamefont {Boucher}}, \ and\ \bibinfo {author} {\bibfnamefont
  {M.}~\bibnamefont {Rubinstein}},\ }\href@noop {} {\bibfield  {journal}
  {\bibinfo  {journal} {Science}\ }\textbf {\bibinfo {volume} {337}},\ \bibinfo
  {pages} {937} (\bibinfo {year} {2012})}\BibitemShut {NoStop}%
\end{thebibliography}%

\end{document}